%% file: ELAA_Channel_Model_Submission.tex
\documentclass[journal]{IEEEtran}
\usepackage{amsmath,amsfonts}
\usepackage{algorithmic,algorithm}
\usepackage{array}
\usepackage{booktabs}
\usepackage{cite}
\usepackage[nonumberlist,acronym,automake,toc,nopostdot]{glossaries}
\usepackage{easyReview}
\usepackage{graphicx}
\usepackage{hyperref}
\usepackage{stfloats}
\usepackage{subfigure}
\usepackage{textcomp}
\usepackage{url}

\makeglossaries
\input{acronyms}
\begin{document}

\title{Near-Field Fading Channel Modeling for ELAAs:\\ From Communication to ISAC}

\author{Jiuyu Liu, Yi Ma, Ahmed Elzanaty, and Rahim Tafazolli
	\thanks{Jiuyu Liu, Yi Ma (corresponding author), Ahmed Elzanaty, and Rahim Tafazolli are with the 5GIC and 6GIC, Institute for Communication Systems (ICS), University of Surrey, Guildford, United Kingdom.}}

\markboth{}%
{Shell \MakeLowercase{\textit{et al.}}: \gls{nf} Statistical Channel Modeling for ELAAs}

\maketitle

\begin{abstract}
\Gls{elaa} is anticipated to serve as a pivotal feature of future \gls{mimo} systems in 6G.
\Gls{nf} fading channel models are essential for reliable link-level simulation and ELAA system design.
In this article, we propose a framework designed to generate \gls{nf} fading channels for both communication and \gls{isac} applications.
The framework allows a mixed of \gls{los} and \gls{nlos} links.
It also considers spherical wave model and spatially non-stationary shadow fading.
Based on this framework, we propose a \gls{3d} fading channel model for ELAA systems deployed with a \gls{ura}.
It can capture the impact of sensing object for ISAC applications.
Moreover, all parameters involved in the framework are based on specifications or measurements from the \gls{3gpp} documents.
Therefore, the proposed framework and channel model have the potential to contribute to the standard in various aspects, including ISAC, extra-large (XL-) MIMO, and \gls{ris} aided MIMO systems.
Finally, future directions for ELAA are presented, including not only NF channel modeling but also the design of next-generation transceivers.

\end{abstract}
\glsresetall

\section{Introduction}
\Gls{elaa} is one of the key candidates for meeting multiple requirements of the \gls{6g} mobile networks, such as massive communications, ubiquitous connectivity, and \gls{isac} \cite{ITUR2023}.
It is a common feature of both \gls{xlmimo} and \gls{ris} aided MIMO systems \cite{Cui2023}.
The Rayleigh distance associated with ELAA often extends beyond hundreds or even thousands of meters \cite{BJORNSON20193}.
This distance significantly greatly exceeds the typical coverage of \glspl{bs} in 5G, situating \glspl{mt} predominantly within the \gls{nf} region of ELAA.
NF-MIMO channels offer extra \gls{dof} compared to the \gls{ff} channels, leading to higher multiplexing gain and spatial resolution \cite{Cui2023}.
However, conventional \gls{ff}-MIMO channel models cannot generalize the unique characteristics of \gls{elaa} channels.
Therefore, accurate and efficient channel models are essential to maximize the benefits that ELAA technology brings to 6G.

\subsection{Overview of Current ELAA Channel Models}
Generally, current ELAA channel models can be mainly divided into three types: \textit{1)} deterministic models, \textit{2)} \glspl{gbsm}, and \textit{3)} fading channel models (see \cite{Liu2023c} and references therein).
For deterministic models, techniques like ray tracing or geometric optics are used to realistically simulate the NF-channel characteristics.
However, their reliance on detailed calculations and intricate path tracing makes them computationally demanding, especially for ELAA systems with a large number of antennas \cite{Zhang2023a}.
In \glspl{gbsm}, clusters are depicted as fundamental units, each representing a set of reflections with similar characteristics.
For each cluster, the channel coefficients are randomly generated based on a certain statistical distribution.
\Glspl{gbsm} demonstrate remarkable adaptability to ISAC applications through the innovative concept of sensing clusters \cite{Zhang2023a}.
Despite their advantages, GBSMs face challenges in their dependence on geometric information, limited suitability, and generality for link-level simulations \cite{Liu2023c}.

In contrast, fading channel models are preferred for link-level simulations in physical layer due to their simple structures \cite{Liu2021}.
The choice of fading model depends on both the propagation condition (i.e., \gls{los} or \gls{nlos} states) and the richness of multipath reflections:
\begin{itemize}
	\item {\bf Rich Reflections in LoS State: } Rician distribution captures the combined effect of \gls{los} and multipath, resulting in a non-central peak.
	Its $K$-factor directly indicates the relative strength of the LoS path.
	\item {\bf Rich Reflections in NLoS State: } Rayleigh distribution effectively captures the behavior of signal power in a diffuse scattering environment. Numerous obstacles randomly redirect the signal before it reaches the receiver.
	\item {\bf Limited Reflections: } Other models such as Nakagami or Weibull distributions have greater flexibility in capturing power imbalances. Depending on the parameter settings, these models can describe the direct LoS path.
\end{itemize}

In FF-MIMO systems, plane-wave model usually supports the assumption that the channel follows \gls{iid} distributions.
However, in NF-MIMO systems, this should be replaced by \gls{swm} to consider distance-dependent pathloss and spherical wavefront \cite{Liu2021}.
By integrating the \gls{swm}, the mentioned fading channel models can be straightforwardly extended to their \gls{ind} versions.
For instance, the i.n.d Rician and Rayleigh distributions are suggested for ELAA in \gls{los} and \gls{nlos} states, respectively \cite{Wang2022,Amiri2018}. 

\begin{figure*}[!t]
	\centering
	\includegraphics[width=0.93\textwidth]{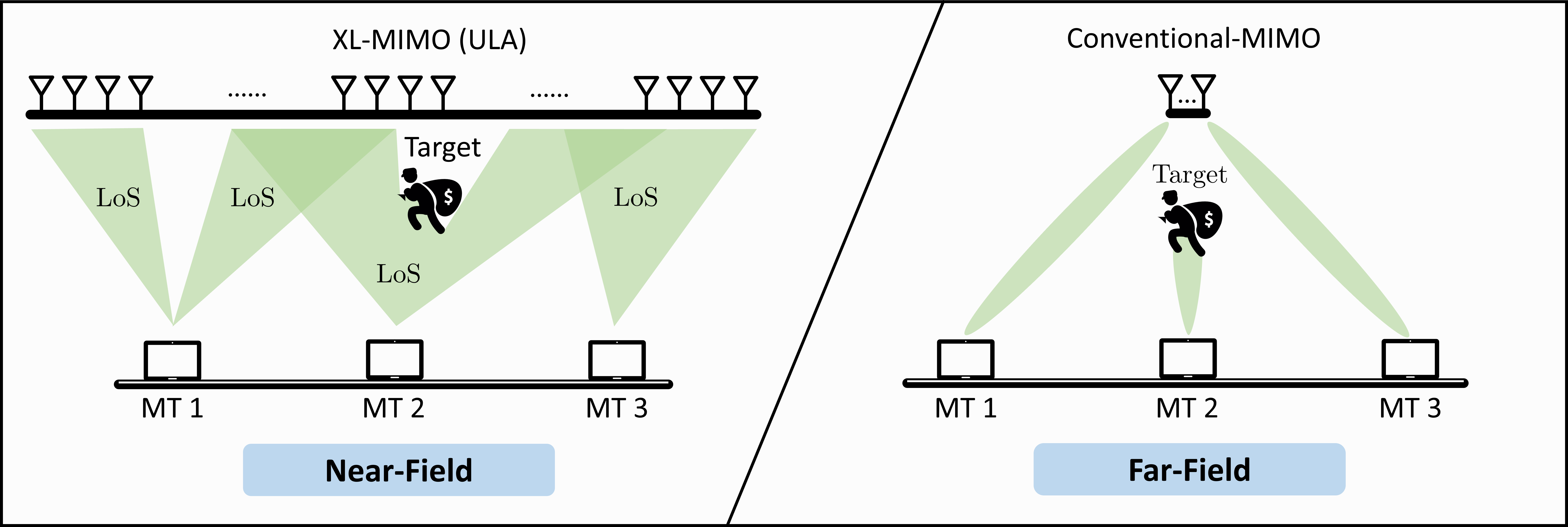}
	\caption{\label{fig_01} A comparison between NF-MIMO and FF-MIMO channels, where the NF-MIMO channel allow a mix of LoS and NLoS links.  The figure demonstrates a typical 3GPP use case for ISAC known as intrusion detection \cite{3gpp.22.837}. In XL-MIMO systems, the sensing target might only obstruct a fraction of the LoS links.}
	\vspace{-1em}
\end{figure*}

Nevertheless, these extensions over-simplify crucial characteristics of ELAA channels \cite{Liu2023c}.
Fig. \ref{fig_01} illustrates a large \gls{ula}, a prevalent form of ELAA.
It is noteworthy that within the NF regions, certain links between a \gls{mt} and ELAA antennas exhibit \gls{los} state, while others exhibit \gls{nlos} state.
This contrasts with FF-MIMO channels, where the entire channel elements between a \gls{mt} and the serving array typically maintains the same \gls{los}/\gls{nlos} state.
The figure demonstrates a use case for ISAC, i.e., intrusion detection, as defined in the \gls{3gpp} \gls{tr} 22.837 \cite{3gpp.22.837}.
In FF-MIMO systems, the sensing target may obstruct the entire LoS links between the MT and \gls{bs}.
However, in XL-MIMO systems, the sensing target might only obstruct a fraction of the LoS links.
Apparently, the mixed of LoS and NLoS links can significantly impact the pathloss and small-scale fading behaviors.
Shadow fading may also exhibit variability across links based on the ELAA channel measurements \cite{Harris2016}.
All these distinctions should be considered in ELAA channel modeling.

To date, the development of accurate fading channel models for ELAA remains an open problem, where the main challenge is to precisely capture the spatially correlated inconsistencies of pathloss, shadow fading and LoS/NLoS states \cite{Cheng2022}.
Recently, a fading channel model has been proposed for the XL-MIMO communication \cite{Liu2023c}.
This model utilizes \gls{edw} to describe the spatial inconsistencies, and can generate computer-simulated channel data that closely matches the field-measurement results \cite{Harris2016}.
This result indicates that a well-designed fading channel model is not only suitable for link-level simulations but also possess the potential to precisely capture NF channel characteristics.
The model is only applicable to XL-MIMO systems using ULA, which indicates that it is \gls{2d} channel model.
However, the more general form of ELAA is represented by \gls{ura}, which can used in both RIS and XL-MIMO, and it requires a \gls{3d} channel model \cite{ETSI2023, Cui2023}.
In addition, sensing capabilities stand out as a crucial requirement for ISAC in 6G \cite{Cui2023, Zhang2023a}.
This highlights the critical importance of modeling general fading channels for ELAA to advance communications and ISAC applications future mobile networks.

\subsection{Contributions of This Article}
This paper provides an extensive review of current literature on channel modeling for ELAA, emphasizing the suitability of fading channel models for link-level simulations.
The review also shows that well-designed fading channel models can accurately capture NF channel characteristics.
So far, the development of NF fading channel models for ELAA is still in its initial stage.
Therefore, the objective of this article is to advance the development of ELAA fading channels to meet the needs of both academic research and industry standards.
The main contributions are summarized as follows:

\begin{enumerate}
	\item At first, we propose a framework for ELAA fading channel modeling applicable to communication and ISAC applications.
	The framework draws inspiration from the in \gls{3gpp} \gls{tr} 38.901 \cite{3gpp.38.901}.
	Moreover, all the parameters involved in this framework are all based on 3GPP specifications or measurements.
	Therefore, the framework can seamlessly integrate with 3GPP standards and make significant contributions to subsequent standards across various areas, including ISAC, XL-MIMO, and RIS-aided MIMO systems.
	\item Given this framework, we then propose a \gls{ura} channel model for communication.
	This model sequentially characterizes the spatial inconsistencies, from horizontal direction to vertical direction.
	Note that a \gls{ula} can be regarded as a specific case of \gls{ura} with only one row.
	Hence, we proposed to use \gls{edw} to characterize the horizontal correlation of the bottom row.
	This allows for the independent generation of vertical correlations that correspond to horizontal windows.
	\item Moreover, a sensing channel is proposed to consider the effect of sensing objects on the communication channel.
	The proposed sensing channel model primarily concentrates on assessing the obstruction caused by the sensing object on the LoS links.
	If the LoS link is blocked, the shadow fading and small-scale fading for the corresponding windows will be regenerated.
	\item Finally, we present future directions of ELAA fading channel modeling and the development of advanced ELAA transceiver technologies for 6G.
	For instance, future ELAA fading channel modeling should consider various other array configurations, such as \gls{uca}, cell-free MIMO, and distributed-MIMO.
\end{enumerate}

\section{The Framework of ELAA Fading Channel Model for Communication and Sensing}
\begin{figure*}[!t]
	\centering
	\includegraphics[width=0.88\textwidth]{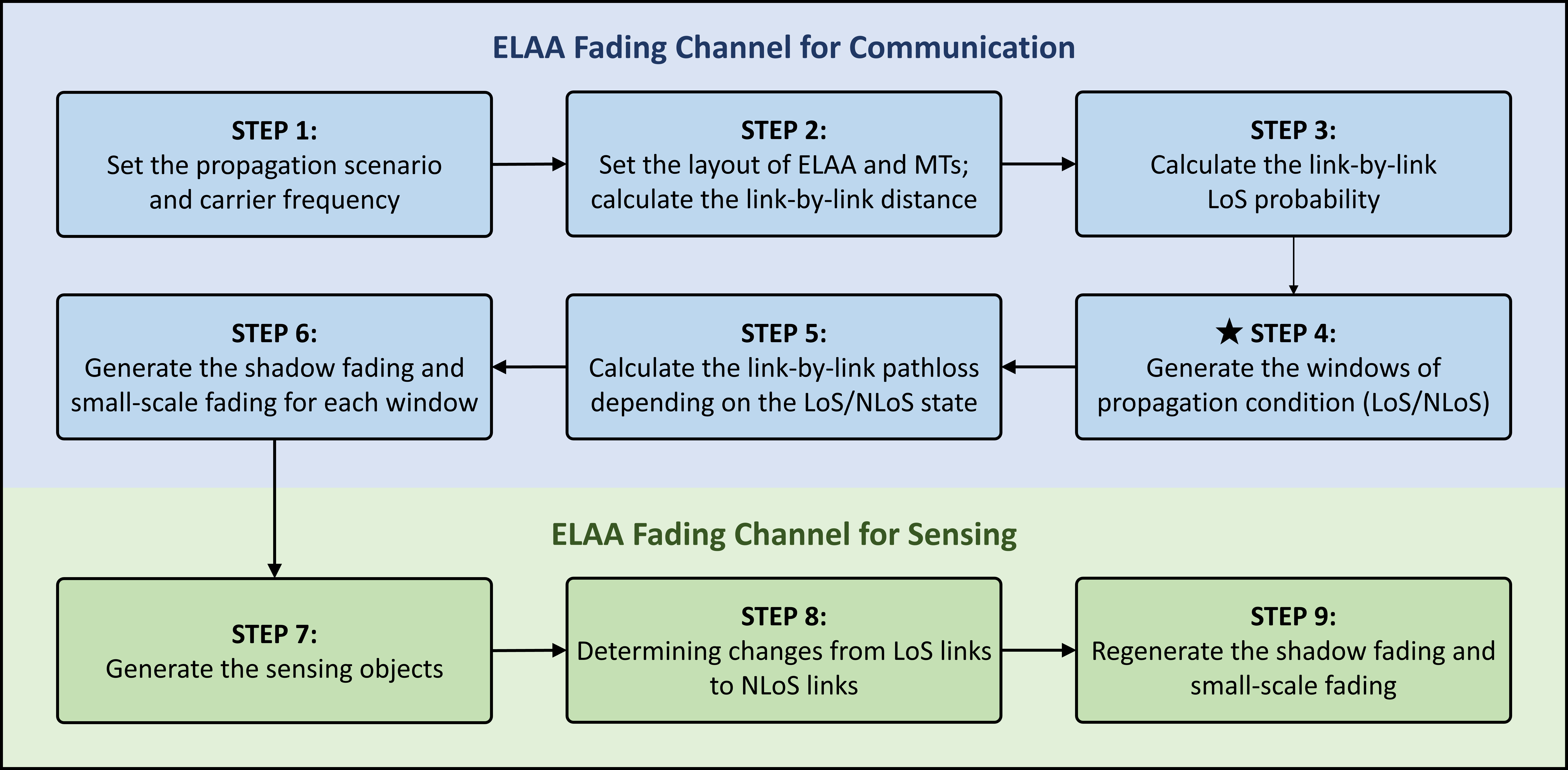}
	\caption{\label{fig_framework} The framework of ELAA fading channel model for both communication and ISAC applications. The proposed framework only requires the coordinates of ELAA, MT and sensing objects, and does not rely on the geometric information of the propagation environment, such as buildings or clusters.}
	\vspace{-1em}
\end{figure*}

In this section, we propose a framework specifically dedicated to modeling ELAA fading channels for both communication and ISAC applications.
This framework draws inspiration from the channel coefficient generation procedure detailed in the \gls{3gpp} \gls{tr} 38.901 \cite[Figure 7.5-1]{3gpp.38.901}.
Note that the 3GPP procedure is part of the GBSM, and its purpose is to generate fast fading coefficients for each cluster based on geometric information.
On the contrary, the proposed framework is designed for generating ELAA fading channels.
This is independent of the specific geometrical details of the environment, such as building layout or furniture arrangement.
As illustrated in Fig. \ref{fig_framework}, the framework encompasses both communication-oriented steps (from {\bf STEP 1} to {\bf STEP 6}) and sensing-oriented steps (from {\bf STEP 7} to {\bf STEP 9}).
The guidelines of the framework are as follows
\begin{itemize}
	\item {\bf STEP 1:} In the 3GPP document, there are six propagation scenarios to choose from, such as \gls{uma}, \gls{umi} and indoor environments \cite{3gpp.38.901}.
	The measurement results show that the \gls{los} probability and pathloss depend on the propagation scenario \cite[Table 7.4.1-1 and 7.4.2-1]{3gpp.38.901}.
	It is also found in the 3GPP standard that the pathloss is related to the carrier frequency.
	\item {\bf STEP 2:} The fading channel model does not rely on detailed environmental geometric information, although it necessitates the layout of ELAA and MT to compute link-to-link distances.
	\item {\bf STEP 3:} Given the distance matrix, the link-by-link LoS probability can be calculated according to the 3GPP measurement results \cite[Table 7.4.2-1]{3gpp.38.901}.
	\item {\bf STEP 4:} This is the most important and challenging step in the framework.
	As discussed before, the ELAA channels allow a mixed of LoS and NLoS links.
	However, it is evident that antennas in close proximity are highly likely to share the same propagation condition.
	In this step, we introduce a concept 'window' to describe this phenomenon.
	A window refers to a set of neighboring ELAA antennas that have identical propagation conditions, while each window is independent.
	The exploration of ELAA window statistics is in its initial stages.
	It has been shown that the EDW can accurately captures the near-field channel characteristics of the ULA \cite{Liu2023c}.
	Moreover, a window generation approach for URA configurations is proposed in this article, as detailed in Section \ref{sec03}.
	Once the windows are generated, we can randomly assign their propagation conditions (LoS/NLoS state) according to the LoS probability, which is calculated in the previous step.
	\item {\bf STEP 5:}	The link-by-link pathloss can be calculated based on the 3GPP measurements results \cite[Table 7.4.1-1]{3gpp.38.901}, where the pathloss models for LoS and NLoS paths depend on the propagation scenario.
	\item {\bf STEP 6:} Shadow fading and small-scale fading can be independently generated for each window.
	Usually, the shadow fading is assumed to follow log-normal distributions. 
	The 3GPP document typically specifies the standard deviations for shadow fading in both \gls{los} and \gls{nlos} states \cite{3gpp.38.901}.
	Furthermore, small-scale fading for \gls{los} and \gls{nlos} states could be assumed to follow Rician and Rayleigh distributions, respectively.
	\item {\bf STEP 7:} This step focuses on creating a point cloud representation of one or more sensing objects. 
	Specific attributes like the number, shape, and precise location of each object are defined within the point cloud.
	\item {\bf STEP 8:} This step aims to determine whether a sensing object obstruct the direct LoS path.
	Given the coordinates of a MT, a set of intersection points in the ELAA plane can be obtained by establishing connections between the MT and all points of the sensing object.
	All of these intersections will form a polygon, and any ELAA elements within the polygon that previously had LoS links will now be blocked by the sensing object.
	This step demonstrates computational feasibility due to its suitability for parallel processing and reliance on relatively simple linear equations with three variables.
	\item {\bf STEP 9:} The sensing object may block certain LoS links of the communication channel generated in the previous six steps.
	In this case, the shadow fading and small-scale fading for the corresponding windows will be regenerated.
\end{itemize}

The proposed framework only requires the coordinates of ELAA, MT and sensing objects, and does not rely on the geometric information of the propagation environment.
Accurately modeling the statistical properties of the window is crucial to ensure that the computer-generated channel data closely aligns with real-world measurements.
Based on this framework, we propose a 3D fading channel model  for URA in the following subsection.

\section{The Proposed 3D Fading Channel Model} \label{sec03}
In this section, we propose a 3D fading channel model designed for communication and sensing applications.
In order to clearly illustrate the main features of the proposed channel model, we will first perform a numerical demonstration.
Following this, we introduce the channel generation process, including communication channels and sensing channels.
The following three subsections are therefore motivated.

\begin{figure*}[!t]
	\centering
	\includegraphics[width=0.92\textwidth]{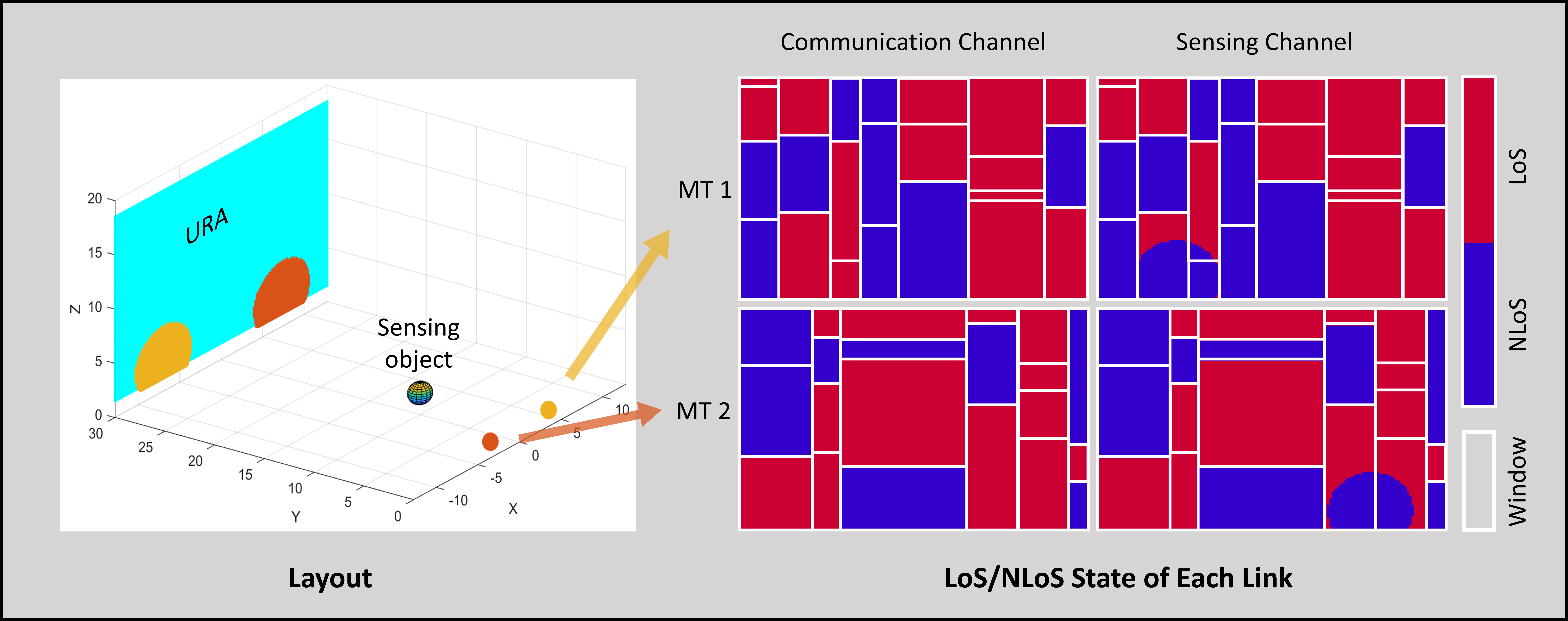}
	\caption{\label{fig_LoS} The numerical demonstration of the proposed fading channel model deploying a URA with 100 × 200  service antennas. There are two MTs and a sphere sensing object. It is obvious that the proposed channel model allows a mixed of LoS and NLoS links.}
	\vspace{-1em}
\end{figure*}

\subsection{The Numerical Demonstration}
\begin{figure*}[!t]
	\centering
	\includegraphics[width=0.60\textwidth]{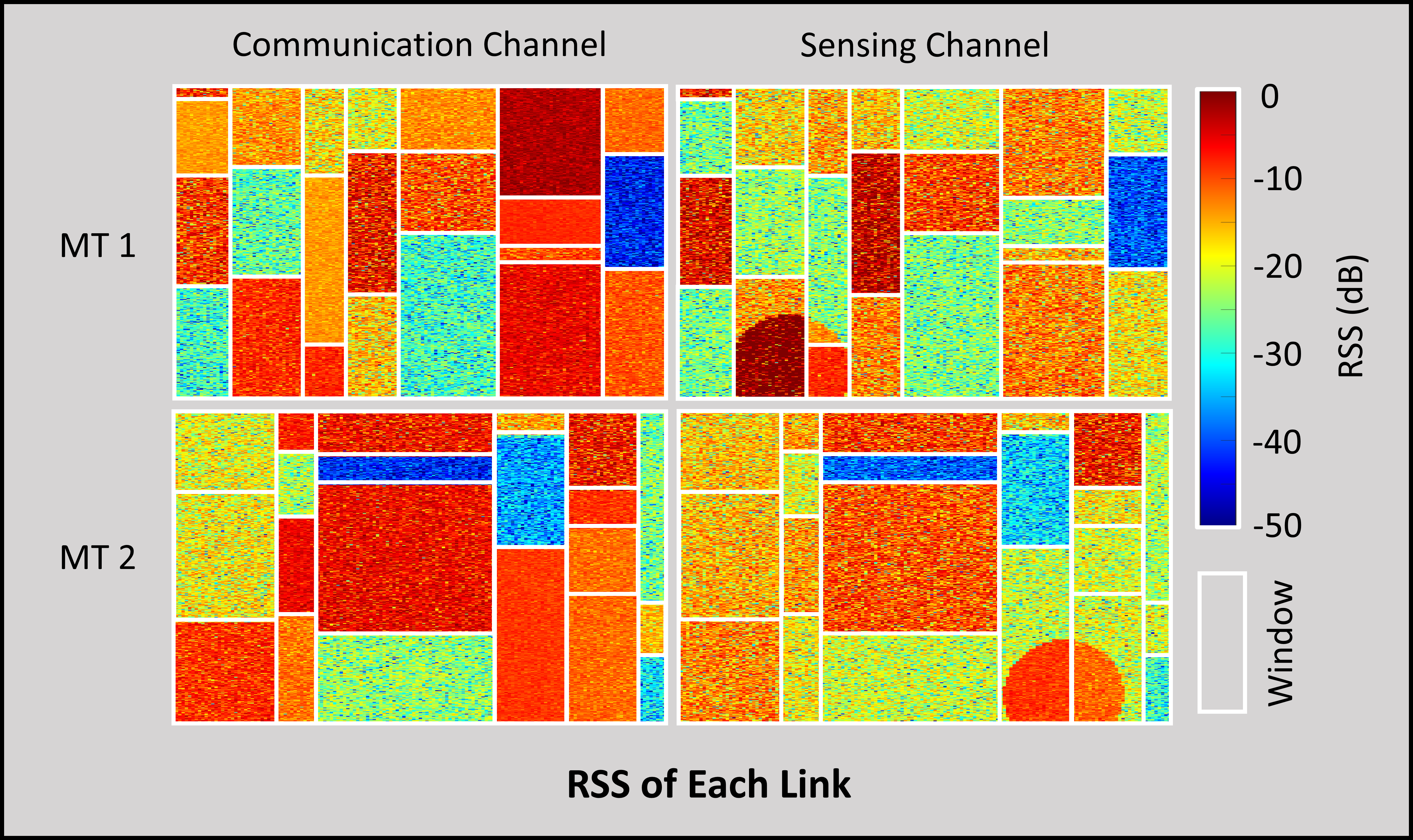}
	\caption{\label{fig_RSS} \gls{rss} of the proposed fading channel model in the same Monte Carlo experiment as Fig. \ref{fig_LoS}. Every element is normalized by the maximum element.}
	\vspace{-1em}
\end{figure*}

As illustrated in Fig. \ref{fig_LoS}, there is a large URA with 100 × 200 antennas and two MTs with 4 antennas each.
For the proposed communication channel, it can be found that the proposed model allows a mixed of LoS and NLoS links.
The URA is randomly divided into several windows (the white rectangles), where the MT-to-URA links share identical propagation condition.
It can be observed that two adjacent windows can have the same propagation condition, but note that their LoS/NLoS states are independently generated.
Moreover, the variation in LoS/NLoS state among different MTs is attributed to their distinct spatial positions.
Furthermore, in Fig. \ref{fig_RSS}, the \gls{rss} of each channel element is shown, normalized by the maximum element.
It is worth noting that the RSS of the LoS window is generally higher than that of the NLoS window.
This difference stems from measurements detailed in the 3GPP document, which state that the pathloss for LoS links is less than for NLoS links at the same distance \cite{3gpp.38.901}.
Moreover, it is observed that neighboring windows within the same LoS/NLoS state may exhibit significantly different RSS. 
This inconsistency can be attributed to shadow fading, which is randomly generated and follows a log-normal distribution according to 3GPP measurement results \cite{3gpp.38.901}.

For the proposed sensing channel, a sensing object (for instance, a sphere) is positioned between the URA and MTs, as depicted in Fig. \ref{fig_LoS}.
As can be seen from the layout, the presence of sensing objects could block the direct LoS path between the MT and the URA.
For instance, the area between MT 1 and the lower left side of the URA, as well as the area between MT 2 and the lower right side of the URA, is clearly obstructed.
In cases where an initial direct LoS path existed within the area during generating the communication channel, all links within that area are transformed into NLoS links.
This phenomenon is appropriately considered in the proposed channel model.
In Fig. \ref{fig_LoS}, it illustrates elliptical NLoS regions in the sensing channels corresponding to MT 1 and MT 2, respectively.
Moreover, when the LoS path encounters an obstruction by the sensing object, the regeneration of both small-scale fading and shadow fading is evident, as explicitly depicted in Fig. \ref{fig_RSS}.

\subsection{The Proposed 3D Fading Channel for Communication}
Modeling the 3D fading channel presents more complexity compared to 2D channel modeling.
For instance, while the \gls{ula} model primarily deals with horizontally linked elements, the \gls{ura} model necessitates consideration for both horizontal and vertical directions.
The main challenge in modeling communication channels is the generation of windows, a process described in {\bf STEP 4} of the framework.
It is important to note that a ULA can be viewed as a particular case of the one-line-only URA.
Therefore, we propose to prioritize the generation of horizontally oriented windows before dealing with vertically oriented windows.

\subsubsection{Window Generation in Horizontal Direction}
In this paper, we propose to use \gls{edw} to characterize the horizontal window of the URA bottom row, which effectively represents a ULA. 
This approach has been shown to accurately capture the ELAA channel characteristics of large ULA \cite{Liu2023c}.
The \gls{pmf} regarding the count of antennas within the same window along the horizontal direction is provided. 
Adjusting a parameter referred to as the correlation distance allows for the modification of its variance.

\subsubsection{Window Generation in Vertical Direction}
Building upon the generated windows for the bottom row, we assume that these windows extend vertically upwards, partitioning the URA into various independent sub-URAs.
This phenomenon is clearly evident in Fig. \ref{fig_LoS} and Fig. \ref{fig_RSS}.
It can be reasonably assumed that antennas in the same row within each sub-URA have identical propagation conditions.
Without loss of generality, let us focus on each sub-URA, and more specifically, a column of the sub-URA.
It is worth noting that a column within the URA essentially constitutes a ULA.
Hence, we suggest utilizing EDW for generating along the vertical direction as well.
Parameters (e.g., correlation distance) may be different in the two directions, but should be verified by future measurements.
Consequently, the entire URA will be randomly segmented into several rectangular windows, each possessing distinct heights and widths.
The propagation conditions for each window can be randomly generated based on the LoS probabilities measured in the 3GPP document \cite{3gpp.38.901}.
If the MT is close to the window, there is a high probability that a LoS path exists.

In the proposed model, it is assumed that the propagation conditions of the two MTs are independently generated. However, the investigation of the correlation between MTs could be a promising future direction of the ELAA channel modeling.
Moreover, it is worth mentioning that future work may consider generating windows with other stochastic processes, such as Markov process or birth-death process \cite{Cheng2022}.

\subsubsection{Pathloss, Small-Scale Fading, and Shadow fading}
In the proposed channel model, the SWM is utilized to compute the pathloss for each link, considering both the distance and the propagation condition.
Subsequently, the small-scale fading and shadow fading is also randomly generated, depending on the LoS/NLoS state of each window.
For instance, if a direct LoS path exists, the small-scale fading of this link can be generated according to Rician distribution.
In addition, the Rician $K$-factor follows log-normal distribution \cite{3gpp.38.901}.
Furthermore, the shadow fading also follows log-normal distribution, but there are distinct standard deviations employed for LoS and NLoS states \cite{3gpp.38.901}.
Channel measurements are essential to validate and potentially calibrate the proposed model for optimal accuracy.
One predictable adjustment relates to shadow fading, since in this article, it is assumed for simplicity that shadow fading shares the same window as the propagation condition.
More accurate modeling of shadow fading is a promising direction for future ELAA fading channel modeling.

\subsection{The Proposed 3D Fading Channel for Sensing}
Within the proposed framework, the sensing object is depicted as a point cloud. Consequently, in computer simulations, the specific shape of the sensing object does not impact the simulation process. 
The spherical sensing object in Fig. \ref{fig_LoS} can be replaced by other shapes, such as a car, a person, or an unmanned aerial vehicle.
Once a sensing object is created, its projected area on the URA plane can be easily determined following {\bf STEP 8} in the proposed framework.
In the blocked area, if direct LoS paths exist for communication channels, all links within the area are converted to NLoS links.
In addition, pathloss, small-scale fading and shadow fading should also be regenerated due to the changed propagation condition.

In summary, unlike current GBSMs and deterministic models reliant on environmental geometric data, the proposed fading channel model solely necessitates the coordinates of the ELAA, the MT, and the sensing objects within 3D space.
Its simple structure is suitable for link-level simulation and has commendable generality.

\section{Discussion: Standards Related Contributions}
The plenary meeting held in December 2023 marked a significant milestone with the conclusion of 3GPP Release 18 (R18), heralding the advent of 5G-Advanced.
The meeting approved the main scope of R19 and outlined a developmental timeline from 5G-Advanced towards 6G, encompassing ISAC, RIS, and the evolution of massive-MIMO \cite{3gpp.22.837, ETSI2023}.
For R19, the ISAC channel modeling study will begin in the second quarter of 2024 for a period of 15 months \cite{3gppWID}.
Moreover, the goal of massive-MIMO in R19 is to support up to 128 \gls{csirs} ports.
This is intended to be consistent with antenna configurations in state-of-the-art massive MIMO deployments, such as XL-MIMO systems.
For this purpose, NF channel modeling is necessary.
However, the existing 3GPP document only cover GBSMs and deterministic models to maintain the accuracy of channel modeling \cite{3gpp.38.901, ETSI2023}.

Recent studies have revealed that even simplistic fading channel models can closely replicate measured channel data, as highlighted in recent research \cite{Liu2023c}. 
This finding suggests the possibility of incorporating accurate NF-MIMO fading channel models into upcoming standards.
The framework proposed in this article takes inspiration from GBSMs outlined in the \gls{3gpp} document \cite{3gpp.38.901}. Moreover, all parameters employed within this framework adhere to 3GPP specifications and measurements. 
All of these features ensure seamless integration and compatibility with existing 3GPP standards.
Consequently, the proposed framework and its associated fading channels hold promising potential to significantly contribute to future standards, potentially impacting R19, R20, and later releases.

\section{Future Directions}
In the following, we mention the technical challenges that are expected to be addressed for ELAA.
The challenges mainly include channel modeling and transceiver design for future NF communication and ISAC applications.

\subsubsection{Modeling the Channel Correlation of Different MTs}
This article assumes independent generation of ELAA channels for different MTs. 
However, in real-world scenarios, closely positioned users tend to exhibit channel correlation, such as in propagation conditions.
Under the assumption of MTs being distributed linearly, their correlation is modeled as linearly associated with the distance between them \cite{Liu2023c}.
Future ELAA channel models may explore diverse MT distributions to accommodate this aspect.

\subsubsection{Modeling the Behavior of NF Shadow Fading}
The assumption that shadow fading and propagation conditions share the same window may not exactly match the actual environment.
While propagation conditions (LoS/NLoS) are binary, shadow fading exists as a continuous variable. 
Hence, it might be more appropriate to depict it with smoother transitions rather than abrupt changes.

\subsubsection{Channel Modeling for Different Types of ELAA}
There is preliminary progress in modeling ELAA fading channels, especially with ULA and URA configurations.
However, there are various ELAA types in 6G, such as UCA, cell-free MIMO, and distributed-MIMO.
This requires additional or more generalized NF-MIMO channel models.

\subsubsection{Field Measurements of ELAA Channels}
The results of the field measurements are valuable data for refining and modifying all proposed theoretical channel models.
Currently, field-measured ELAA channels are primarily limited to ULA configurations and operate in sub-6 GHz carrier frequencies.
Field measurements of different types of ELAA and higher carrier frequencies are essential for ELAA techniques.
For instance, 3GPP R19 specifies that future 6G systems will use frequencies in the 7-24 GHz range, but there is a lack of measurements in this frequency range.

\subsubsection{ELAA Channel Estimation}
The NF effect poses challenges to channel estimation.
Taking \gls{lmmse} channel estimator as an example, it requires a Gram matrix when estimating the MIMO channel.
In FF-MIMO systems, the Gram matrix can be approximated as the identity matrix.
Nevertheless, due to the NF effects, determining the precise Gram matrix utilized for ELAA channel estimation is not feasible \cite{Lu2023}.
The development of estimation algorithms capable of adapting to the dynamic nature of NF-MIMO channels is crucial.
Also, exploiting the inherent sparsity of NF-MIMO channels can be a powerful strategy for reducing estimation complexity and improving performance.

\subsubsection{Transceivers Design for XL-MIMO Systems}
ELAA is expected to achieve unprecedented spectral efficiency and full multiplexing gains in the future to support massive MTs.
This vision depends on complex spatial multiplexing techniques, such as \gls{zf} and \gls{lmmse}.
However, these techniques entail cubic order complexity.
While the iterative algorithm exhibits lower quadratic complexity, it suffers from sluggish convergence in ill-conditioned ELAA channels, negating its complexity advantage.
There is an urgent need to discover a superior solution for this challenge.

\subsubsection{Algorithm Design for NF Sensing and Localization}
ELAA channels can offer unprecedented \gls{dof} for sensing and localization applications.
To fully utilize this potential, existing algorithms must be redesigned to match the unique characteristics of the ELAA channel.
For instance, design algorithms that effectively exploit the full 3D spatial information available in ELAA channels, enabling more precise localization and fine-grained sensing capabilities.
Additionally, develop algorithms that meticulously model and leverage the spatial correlations inherent in NF-MIMO channels, leading to enhanced accuracy and robustness in both sensing and localization tasks.

\section{Conclusion}
In this article, we propose a novel framework for generating ELAA fading channels, covering both communication and sensing channels.
The proposed framework captures crucial NF effects, incorporating spatial inconsistencies of pathloss, shadow fading, and propagation conditions.
We then present a 3D fading channel model tailored for URA configurations, building upon the existing state-of-the-art 2D communication channel designed for ULA.
Notably, our model captures the influence of sensing objects, specifically addressing ISAC applications. All parameters utilized in this framework are carefully chosen from 3GPP specifications and measurements.
Therefore, our proposed framework and channel model exhibit significant promise for adoption in upcoming standards involving ISAC, XL-MIMO, and RIS-aided MIMO technologies. 
Beyond channel modeling, the future of ELAA encompasses innovative transceiver designs, crucial for its overall success.

\section*{Acknowledgment}
This work was partially funded by the 5G and 6G Innovation Centre, University of Surrey, and partially by the UK national TUDOR project.

\ifCLASSOPTIONcaptionsoff
\newpage
\fi

\bibliographystyle{myIEEEtran}
\bibliography{IEEEabrv,mMIMO} 

\section*{Biographies}

\vspace{-1em}
~\\
\textsc{Jiuyu Liu} (jiuyu.liu@surrey.ac.uk) is currently pursuing his Ph.D. degree in electronic engineering with the 5GIC \& 6GIC, Institute for Communication Systems, University of Surrey, U.K.
He received the M.Sc. degree in electronic engineering from the University
of Surrey, U.K., in 2020.
His main research interests include multiple-input multiple-output, extremely large aperture array, near-field channel modeling, and digital signal processing.

~\\
\textsc{Yi Ma} (y.ma@surrey.ac.uk) is a Chair Professor within the Institute for Communication Systems (ICS), University of Surrey, Guildford, U.K.
He has authored or co-authored 200+ peer-reviewed IEEE journals and conference papers in the areas of deep learning, cooperative communications, cognitive radios, interference utilization, cooperative localization, radio resource allocation, multiple-input multiple-output, estimation, synchronization, and modulation and detection techniques.
He holds 10 international patents in the areas of spectrum sensing and signal modulation and detection.
He has served as the Tutorial Chair for EuroWireless2013, PIMRC2014, and CAMAD2015. He was the Founder of the Crowd-Net Workshop in conjunction with ICC’15, ICC’16, and ICC’17. 
He is the Co-Chair of the Signal Processing for Communications Symposium in ICC’19.

~\\
\textsc{Ahmed Elzanaty} (a.elzanaty@surrey.ac.uk) is currently a Lecturer (an Assistant Professor) with the Institute for Communication Systems, University of Surrey, U.K. 
He received the Ph.D. degree (cum laude) in electronics, telecommunications, and information technology from the University of Bologna, Italy, in 2018.
He was a Post-Doctoral Fellow with the King Abdullah University of Science and Technology (KAUST), Saudi Arabia. 
He has participated in several national and European projects, such as TUDOR, GRETA, and EuroCPS. 
His research interests include the design and performance analysis of wireless communications and localization systems.

~\\
\textsc{Rahim Tafazolli} (r.tafazolli@surrey.ac.uk) is Regius Professor of Electronic Engineering, Professor of Mobile and Satellite Communications, Founder and Director of 5GIC, 6GIC and ICS (Institute for Communication System) at the University of Surrey.
He has authored and co-authored 1,000+ research publications and is regularly invited to deliver keynote talks and distinguished lectures to international conferences and workshops.
He was the leader of study on “grand challenges in IoT” (Internet of Things) in the UK, 2011-2012, for RCUK (Research Council UK) and the UK TSB (Technology Strategy Board).
He is the Editor of two books on Technologies for Wireless Future (Wiley) vol.1, in 2004 and vol. 2, in 2006. 
He holds Fellowship of Royal Academy of Engineering (FREng), Institute of Engineering and Technology (FIET) as well as that of Wireless World Research Forum. 
He was also awarded the 28th KIA Laureate Award- 2015 for his contribution to communications technology.

\end{document}

%% file: Acronyms.tex
\newacronym{1d}{1D}{one-dimensional}
\newacronym{2d}{2D}{two-dimensional }
\newacronym{3d}{3D}{three-dimensional}
\newacronym{3gpp}{3GPP}{3rd Generation Partnership Project}
\newacronym{5g}{5G}{fifth-generation}
\newacronym{6g}{6G}{sixth-generation}

\newacronym{ask}{ASK}{amplitude shift keying}
\newacronym{awgn}{AWGN}{additive white Gaussian noise}

\newacronym{ber}{BER}{bit error rate}
\newacronym{bler}{BLER}{bit error rate}
\newacronym{bp}{BP}{belief propagation}
\newacronym{bpsk}{BPSK}{binary phase shift keying}
\newacronym{bs}{BS}{base station}

\newacronym{cbsm}{CBSM}{correlation-based stochastic model}
\newacronym{csirs}{CSI-RS}{channel state information reference signal}
\newacronym{cdf}{CDF}{cumulative distribution function}
\newacronym{cdma}{CDMA}{code division multiple access}
\newacronym{clt}{CLT}{central limit theorem}
\newacronym{csi}{CSI}{Channel state information }
\newacronym{cvp}{CVP}{closest vector problem}

\newacronym{dof}{DoF}{degrees of freedom}

\newacronym{edw}{EDW}{exponentially decaying window}
\newacronym{elaa}{ELAA}{extremely large aperture array}
\newacronym{etsi}{ETSI}{European Telecommunications Standards Institute}

\newacronym{ff}{FF}{far-field}
\newacronym{fcsd}{FCSD}{fixed-complexity sphere decoder}
\newacronym{fec}{FEC}{forward error correction}
\newacronym{fspl}{FSPL}{free space path loss}

\newacronym{gbsm}{GBSM}{geometry-based stochastic model}
\newacronym{gmsk}{GMSK}{Gaussian minimum shift keying}
\newacronym{gsm}{GSM}{global system for mobile communication}

\newacronym{iid}{i.i.d.}{independently and identical distributed}
\newacronym{ind}{i.n.d.}{independently and non-identical distributed}
\newacronym{imt}{IMT}{International Mobile Telecommunications}
\newacronym{isac}{ISAC}{integrated sensing and communication}
\newacronym{itur}{ITU-R}{International Telecommunication Union Radiocommunication Sector}

\newacronym{jsac}{JSAC}{joint sensing and communication}

\newacronym{las}{LAS}{likelihood ascent search}
\newacronym{lll}{LLL}{Lenstra-Lenstra-Lov\'{a}sz}
\newacronym{llr}{LLR}{log-likelihood ratio}
\newacronym{los}{LoS}{line of sight}
\newacronym{lr}{LR}{lattice reduction}
\newacronym{lmmse}{LMMSE}{linear minimum mean square error}
\newacronym{lsd}{LSD}{list sphere decoder}
\newacronym{lte}{LTE}{long-term evolution }

\newacronym{map}{MAP}{maximum a posteriori}
\newacronym{mf}{MF}{matched filter}
\newacronym{mimo}{MIMO}{multiple-input multiple-output}
\newacronym{ml}{ML}{maximum likelihood}
\newacronym{mse}{MSE}{mean square error}
\newacronym{mt}{MT}{mobile terminal}
\newacronym{nf}{NF}{near-field}
\newacronym{nlos}{NLoS}{non-LoS}

\newacronym{od}{OD}{orthogonality defect}

\newacronym{pdf}{PDF}{probability distribution function}
\newacronym{pda}{PDA}{probabilistic data association}
\newacronym{pep}{PEP}{pairwise error probability}
\newacronym{pl}{PL}{path-loss}
\newacronym{pmf}{PMF}{probability mass function}
\newacronym{pwm}{PWM}{plane-wave model}

\newacronym{qam}{QAM}{quadrature amplitude modulation}
\newacronym{qpsk}{QPSK}{quadrature phase shift keying}

\newacronym{ris}{RIS}{reconfigurable intelligent surface}
\newacronym{rss}{RSS}{received signal strength}

\newacronym{sa}{SA}{Seysen's algorithm}
\newacronym{sd}{SD}{sphere decoder}
\newacronym{sdr}{SDR}{semidefinite relaxation}
\newacronym{ser}{SER}{symbol error rate}
\newacronym{sf}{SF}{shadow fading}
\newacronym{sic}{SIC}{succesive interference cancellation}
\newacronym{sinr}{SINR}{signal-to-interference-plus-noise ratio}
\newacronym{siso}{SISO}{single input single output}
\newacronym{snr}{SNR}{signal-to-noise ratio}
\newacronym{sns}{SNS}{spatial non-stationarity}
\newacronym{sota}{SoTA}{state-of-the-art}
\newacronym{svd}{SVD}{Singular value decomposition }
\newacronym{swm}{SWM}{spherical-wave model}

\newacronym{tr}{TR}{Technical Report}
\newacronym{ts}{TS}{tabu search}

\newacronym{uca}{UCA}{uniform cylindrical array}
\newacronym{ue}{UE}{user equipment}
\newacronym{ula}{ULA}{uniform linear array}
\newacronym{ura}{URA}{uniform rectangular array}
\newacronym{uma}{UMa}{urban macro}
\newacronym{umi}{UMi}{urban micro}
\newacronym{upa}{UPA}{uniform planar array}

\newacronym{vblast}{V-BLAST}{vertical Bell Labs layered space-time}

\newacronym{xlmimo}{XL-MIMO}{extra-large multiple-input multiple-output}

\newacronym{zf}{ZF}{zero forcing}